\documentclass[tighten, twocolumn, trackchanges]{aastex699}	
\usepackage{xcolor, fontawesome, microtype, amsmath}	
\definecolor{twitterblue}{RGB}{64,153,255}	
\definecolor{linkcolor}{rgb}{0.1216,0.4667,0.7059}	
\newcommand{\twitter}[1]{\href{https://twitter.com/#1}{\textcolor{twitterblue}{\faTwitter}\,\tt \textcolor{twitterblue}{@#1}}}	
\shorttitle{Interstellar communication network. III. Locating deep space nodes}	
\shortauthors{Michael Hippke}	
\begin{document}	
\title{Interstellar communication network. III. Locating deep space nodes}	
\author[0000-0002-0794-6339]{Michael Hippke}	
\affiliation{Sonneberg Observatory, Sternwartestr. 32, 96515 Sonneberg, Germany \twitter{hippke}}
\affiliation{Visiting Scholar, Breakthrough Listen Group, Berkeley SETI Research Center, Astronomy Department, UC Berkeley}
\email{michael@hippke.org}

\begin{abstract}	
An interstellar communication network benefits from relay nodes placed in the gravitational lenses of stars. The signal gains are of order $10^{9}$ with optimal alignment, allowing for GBits connections at kW power levels with meter-sized probes over parsec distances. If such a network exists, there might be a node in our solar system: where is it? With some assumptions on the network topology, candidate sky positions can be calculated. Apparent positions are influenced by the parallax motion from the Earth's orbit around the Sun, and the (slow) drifts caused by proper motions of nearby stars. With Gaia astrometry, instantaneous positions can be determined with arcsec accuracy. These potential node locations can be observed in targeted SETI experiments.	
\end{abstract}	


\section{Introduction}	
\begin{figure*}	
\includegraphics[width=\linewidth]{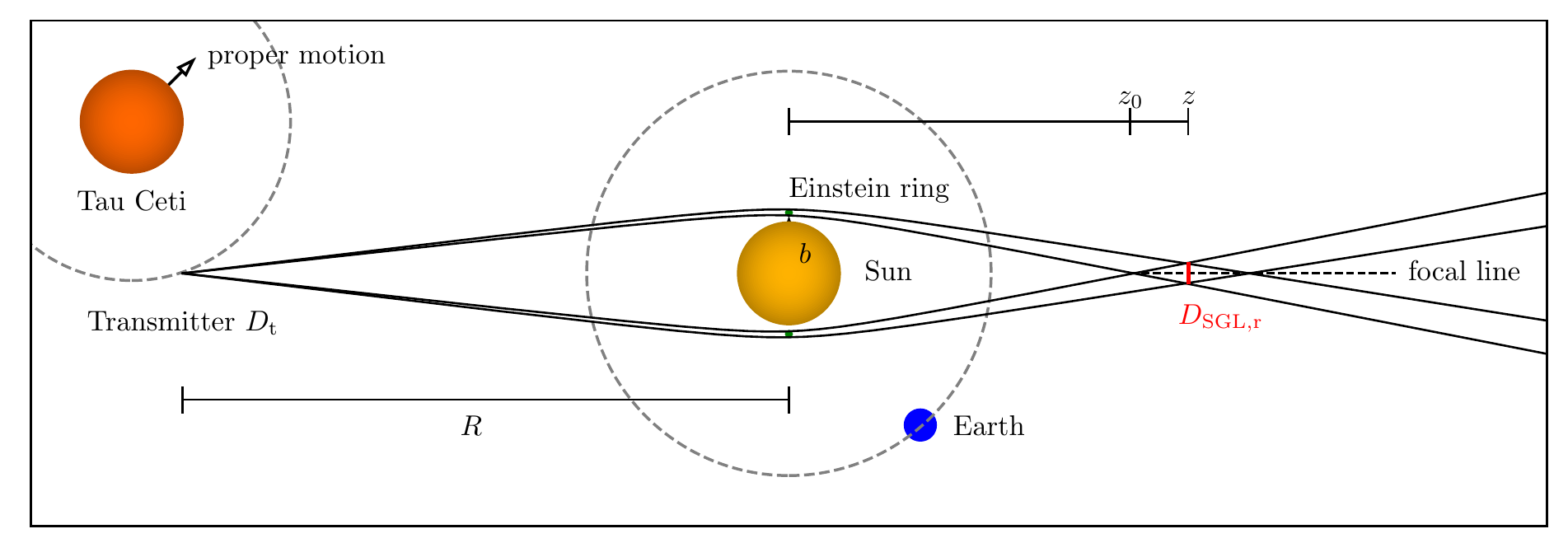}	
\caption{Cartoon diagram of the solar gravitational lens configuration. The receiver with aperture $D_{\rm SGL,r}$ on the focal line (right) observes the flux at distance $z$ from the sun which comes through the Einstein ring from the transmitter at distance $d$.}
\label{fig:ray}	
\end{figure*}

In the second part of this series \citep{2020arXiv200901866H} it was shown that an interstellar network \citep{2019arXiv191202616H} benefits greatly from the gain provided by stellar (solar) gravitational lensing (SGL). One of the nodes in a link must be located in the image plane \citep{Landis2017,2018JBIS...71..369L}, where a receiver can achieve a gain of order $10^9$. The physical sizes of nodes are expected to be about a meter, positioned of order $1{,}000$ astronomical units from the sun.	
This interstellar communication scheme (Figure~\ref{fig:ray}) is the first concrete proposal for such a system. It includes physical constraints on the size and placement of nodes, and the characteristics of the links including power requirements, wavelength, and data rate. The scheme is superior to uncoordinated communications on the direct path with respect to data rates and energy efficiency. 

Yet, we can not be sure that the design is the best possible. With further research and technological progress, we may find ideas to improve it. Proposals to do so are a valuable step towards convergence onto the best system. In the future, humans may then build such a system. Similarly, other species might have built one in the deep past. If a galactic network exists, we should strive to describe its physics to infer the locations of nodes.

In addition to communication nodes best located in deep space, there might also exist exploration probes in the inner solar system, closer to the object of interest. Explorers would gather intelligence and surveillance, to be transmitted via connections to deep space nodes. The next paper in the series will cover this tangent.	
The present paper covers the question of how to locate deep space nodes in the gravitational lens of our sun. 

Our lensing network concept is the first to predict exact locations of alien probes. To locate a receiving node in the image plane, three steps need to be completed. First, we need to guess the origin of the transmission from which the probe receives data. For example, we could assume that there exists a network line from Tau Ceti to our solar system. A prioritized list of such guesses is discussed in section~\ref{sec:list}. Second, we need the location of the probe in the heliocentric reference frame. As argued in \citet{2020arXiv200901866H}, it is most likely at a heliocentric distance $1000<z<2{,}500\,$au. Third, given a combination of the former, we can calculate the apparent location of this node, as seen from the Earth, and estimate the uncertainties (section~\ref{sec:localization_receiver}). Similar to this process, step three can be adjusted to search for transmitters in the SGL (section~\ref{sec:localization_transmitter}).	
We provide an open source online calculator\footnote{\url{https://github.com/hippke/sgl}} to determine the node position in our solar system, as a function of link partner and observation time. The tool can be used to reproduce the figures of the paper, and to plan SETA (search for extraterrestrial artifacts, section~\ref{sec:seta}) and META (messaging to ETA, section~\ref{sec:meta}) observations.	

\section{Priority target list}	
\label{sec:list}	
The communication scheme assumes a mixture of exploration probes and data relay nodes across star systems. As argued in \citet{2019arXiv191202616H}, distances between nodes are likely a few parsec. It may be that probes are very common. Alternatively, one could assume that planets with biological features attract probes and encourage the installation of deep space nodes in such systems. As of now, no exoplanets with life are known. In the next decades, discoveries may become possible with transit spectroscopy, which can reveal biosignatures in atmospheres \citep{2014PNAS..11112634S,2018AJ....156..114K}. In the meantime, we can only assume that some nearby star systems host a probe, which connects to a node in our solar system. Thus, the initial list of candidate stars with nodes should include the nearest stars (Alpha and Proxima Cen), the nearest solitary G-dwarfs \citep[e.g., Tau Ceti,][]{mccollum1992sails}. 	
The list could be extended to stars with known and interesting exoplanets such \mbox{TRAPPIST-1} \citep{2017Natur.542..456G}. Larger lists could leverage catalogs designed for similar purposes like HabCat \citep{2003ApJS..145..181T,2003ApJS..149..423T}. More exotic targets could also be included, because there is a large uncertainty about life as we do \textit{not} know it \citep{2020arXiv200100673G}. Humans would certainly be interested in sensor data of probes around astrophysically interesting objects such as pulsars and hypergiant stars. Candidates could include the nearest and shortest-period pulsars \citep{2017arXiv171106036V,2017arXiv170403316V} and the galactic center \citep{2011arXiv1104.4362V,2011CQGra..28w5015D,2012GrCo...18...65D,2017AmJPh..85...14O,2019arXiv190310698A} and anti-center \citep{2018ApJ...856...31T}. Finally, brown dwarf atmospheres have been argued to offer the largest chunk of habitable real estate \citep{2019ApJ...883..143L}.

The lensing gain is of order $10^9$ for all main-sequence stars, and varies only within a factor of a few, with more massive stars being advantageous. Supergiants are impractical, because their large radii cause the minimum focus to be very far out, and their high luminosity is a major source of noise. In contrast, small, massive, dark bodies are optimal. The lensing gain for the galactic black hole is about three orders of magnitude larger than for the sun. It is also a strong ``Schelling point'' \citep{2020IJAsB..19..515W}. Double stars are problematic because they distort the gravitational field and require complex course corrections. To a lesser extent, systems with high-mass planets are similarly effected, although the effect on the resulting data rate is yet to be determined. 

Conservatively, we argue that the most likely routes are to and from the nearest systems, and out of these, the nearest solitary G-dwarfs. After all, habitability of planets around the more numerous M-dwarfs is doubtful \citep{2016PhR...663....1S}. A first attempt for a prioritized list is shown in Table~\ref{tab:targets}, which can be refined with future knowledge, and be used to start observations.	
\begin{table*}	
\center	
\caption{Nearest network node: location priority list (top 10)	
\label{tab:targets}}	
\begin{tabular}{lccl}	
\tableline	
Object          & Spectral Type & Distance (pc) & Rationale \\	
\tableline	
Proxima Cen     & M5.5Ve     & 1.3 & closest star; rocky planet \citep{2016Natur.536..437A} \\	
Tau Ceti & G8V  & 3.6 & nearest solitary G-dwarf, planets? \citep{2017AJ....154..135F} \\	
Sigma Draconis  & G9V & 5.6 & 2nd closest G-dwarf \\	
Epsilon Eridani & G6V & 3.2 & circumstellar disks; planets? \citep{2008AA...488..771J} \\	
Delta Pavonis   & G8IV & 6.0 & ``best SETI target'' \citep{2003ApJS..149..423T} \\	
Barnard's Star  & M4Ve & 1.8 & largest proper motion, planet \citep{2018Natur.563..365R} \\	
Wolf 359        & M6V  & 2.4 & 2 planets \citep{2019arXiv190604644T} \\	
Lalande 21185   & M2V & 2.5 & 1 planet \citep{2019AA...625A..17D} \\
$\alpha$ Cen AB & G2V \& K1V & 1.3 & closest G-dwarf, but binary \\
Sirius AB       & A1V \& DA2 & 2.6& brightest star in the sky, but binary \\	
Sgr A$^{\star}$ & black hole & 8{,}178 & Schelling point \\
\tableline
\end{tabular}	
\end{table*}

\begin{figure*}
\includegraphics[width=\linewidth]{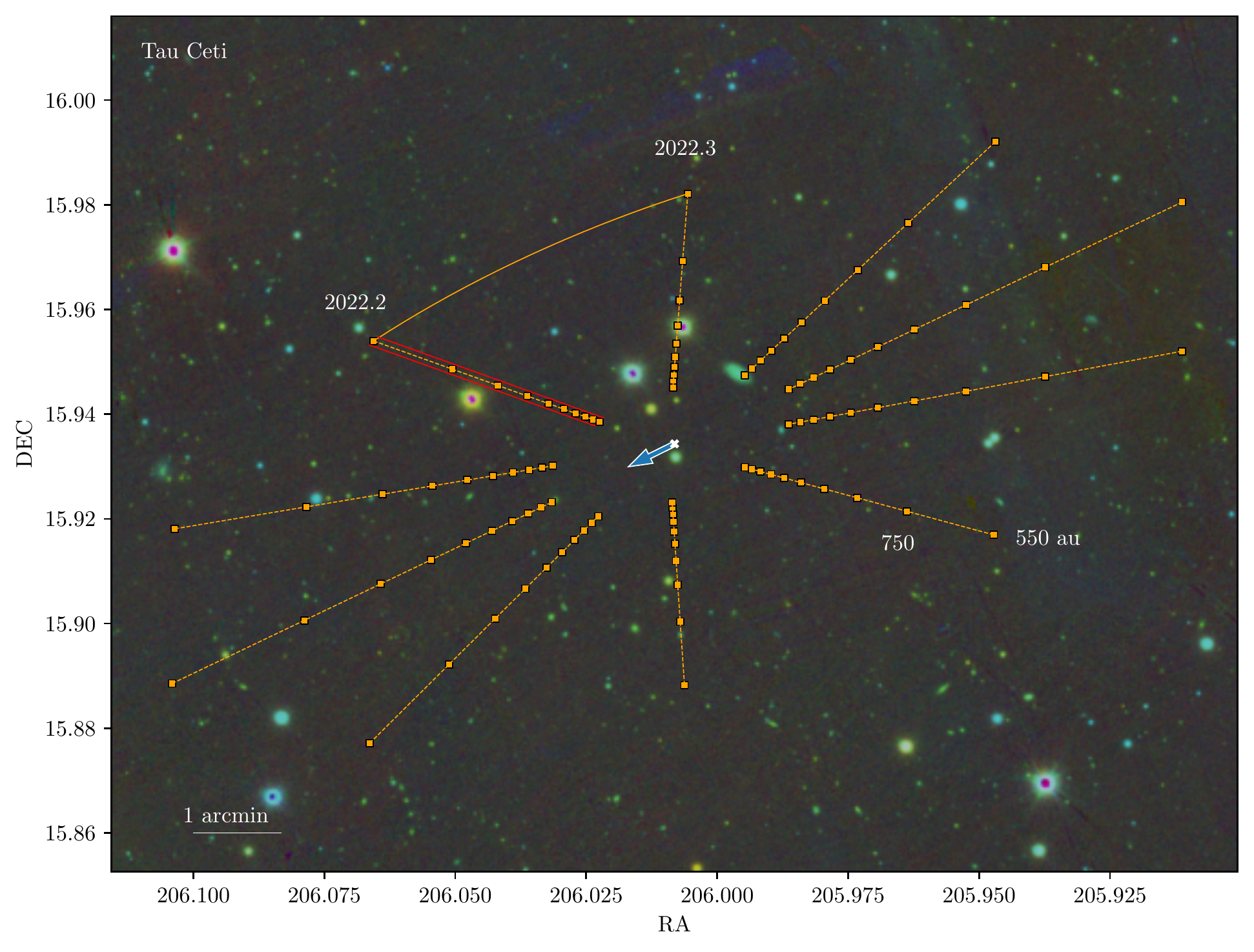}
\caption{Location of a probe in the sun's gravitational lens for Tau Ceti, as seen from the Earth. The parallax due to Earth's orbit around the sun over the course of a year translates into an ellipsoid path. The outer line represents such an apparent probe movement between 2020.2 and 2020.3, for a heliocentric distance of 550\,au. At greater distances (e.g., 750\,au), the resulting ellipse becomes smaller (square markers in steps of 200\,au). The arrow in the center shows the shift of the ellipses due to Tau Ceti's proper motion over the next 10 years. The two red lines flanking the 2020.2 epoch give the apparent location shift due to a transmitter with a 10\,au distance from Tau Ceti. The sky image is the {\it grizy} sum from Pan-STARRS \citep{2016arXiv161205560C,2016arXiv161205243F} taken in 2015.}
\label{fig:tau_ceti_large}
\end{figure*}

\section{Localization of a receiver node in the solar gravitational lens}
\label{sec:localization_receiver}
The alignment between receivers and transmitters, as well as the physical constraints on their aperture sizes, wavelengths, and powers, was given in detail in \citet{2020arXiv200901866H}. Briefly, in the basic scheme a transmitter at pc distance beams at the Sun, taking into account proper motion. The beam width at arrival is at least a few solar radii, and the gravitational field deflects the incoming light rays into a focal line starting at 550\,au behind the Sun. A receiver with aperture $D_{\rm SGL, r}$ in the image plane collects part of this flux with a gain of order $10^9$ compared to the direct path. In the reversed scheme, the probe in the image plane becomes a transmitter with aperture $D_{\rm SGL, t}$ and beams part of its flux into the Einstein ring or an arc segment. 

The location of the SGL, as seen from the Earth, was calculated by \citet{2014AcAau..94..629G} using Alpha Centauri as an example. The computation, however, was simplified because it marginalized over Earth's yearly orbit around the sun. The simple procedure results in a sky area in the shape of an ellipse with a semimajor axis of $\sim10$ arcmin.

In this paper, we calculate the SGL position inside this ellipse for a moment of time (when the observation is taken) as seen from the Earth. Over the course of a year (one Earth orbit around the Sun), the apparent SGL position circles around inside this ellipse. The maximum diameter of the ellipse is defined by the probe's (unknown) heliocentric distance $z>550\,$au, with distances closer to the Earth appearing further away from the center. For a specific point in time, the probe's position is a point on a line which covers a range of $z$ values.

To get to the arcsecond level, we need to calculate the exact positions of all bodies involved, and their constant motions. We follow the path of photons from a distant transmitter to a receiver in the SGL. There are five bodies to consider: 
The transmitter near a distant star, 
the sun, 
an observer on Earth, 
and the probe in the SGL 
(Figure~\ref{fig:ray}).

\subsection{Vector calculation}
We make the calculation using a specific example of a transmitter near Tau Ceti \citep{niven1987gift}, the closest solitary G-class star (G8V, $0.8\,R_{\odot}$) at a distance of 3.65\,pc. The star might be host to five or more planets, of which two could be in the habitable zone \citep{2013A&A...551A..79T,2017AJ....154..135F}, as well as a debris disk \citep{2016ApJ...828..113M}. 

We assume that the transmitter is on a circular orbit around Tau Ceti with a semi-major axis of 1\,au, but unknown inclination. Later, we will determine the effect of different transmitter separations.

Astrometry from Gaia DR2 \citep{ 2018A&A...616A...1G} at epoch 2015.5 provides 
${\rm RA}=26.0093028(1)$, 
${\rm DEC}=-15.93379865(9)$,  
with a parallax of $277.5\pm0.5\,$mas, 
i.e. a distance of $3.6036\pm0.0065\,$pc ($\widehat{=} \pm 1341\,$au).
The proper motions are 
$-1729.7\pm1.3\,$mas\,yr$^{-1}$ in RA and
$855.5\pm0.8\,$mas\,yr$^{-1}$ in DEC.

We estimate the location of the SGL for an example epoch of 2020.2, using the location, proper motions, and corresponding uncertainties. A first correction must be made due to finite light speed. As we will see in the following, the apparent position in the sky changes fast enough (due to the parallax motion of the Earth around the sun) so that the light travel time has to be taken into account. With a light travel time of 499\,s per au, we see the sky at $1{,}000\,$au as it was 5.8 days ago -- shifted by several arcsec. As this is a function of heliocentric distance $z$, the epoch has to be modified for each case of $z$ individually. We do this by subtracting the light travel time from the epoch before calculating the vectors. Then, we see the probe's location at observation data 2020.2 where it was at $2020.2-0.0158=2020.1841$.

At this epoch, the location of Tau Ceti is
${\rm RA}=26.007(2)$ and
${\rm DEC}=-15.9326(1)$, 
when propagating the Gaia uncertainties over 4.7\,yrs, i.e. errors of $\lesssim1\,$arcsec. Photons sent from a transmitter close to Tau Ceti in the past will appear near this location at 2020.1841, where we observe the probe at 2020.2. The probe travels an apparent distance of 19\,arcsec in these 5.8\,days.

We transform the spherical coordinates into cartesian coordinates using the {\tt astropy.coordinates} routine \citep{2013A&A...558A..33A,2018AJ....156..123A} where the Earth is at $ X=Y=Z=0$. For the sun, we obtain the location in the same coordinate system using the JPL DE430 ephemerides \citep{1994A&A...287..279F}:

\begin{equation}
{\rm S} = 
 \begin{pmatrix}
   0.97103 \\
  -0.18896 \\
  -0.08191 \\
 \end{pmatrix}
 {\rm au.}
\end{equation}

For Tau Ceti:

\begin{equation}
 {\rm T} = 
 \begin{pmatrix}
   642314 \pm 1160  \\
   313406 \pm 566 \\
  -204042 \pm 368 \\
 \end{pmatrix}
 {\rm au.}
\end{equation}

We denote the vector from the Sun to Tau Ceti as $\overrightarrow{ST}$, and its length as $\lvert ST \rvert$. To reach the SGL, we start from the Sun along the path of this vector, backwards by $z$ (see Figure~\ref{fig:ray}). Thus, we calculate the location of the SGL as

\begin{equation}
 {\rm P} = S - \overrightarrow{ST} \cdot \frac{z}{\lvert ST \rvert} = 
 \begin{pmatrix}
   -863.2220(2) \\
   -421.8587(1) \\
    274.44476(1) \\
 \end{pmatrix}
 {\rm au}
\end{equation}

from the coordinate origin. The uncertainty of the probe location is of order $10{,}000\,$km in the image plane, consistent with the plate scale of $z/d$. We translate back into spherical coordinates
RA=206.044871(6),
DEC=15.941723(3) 
as seen from Greenwich, Earth using {\tt astropy EarthLocation}. The difference from using any other observatory location is negligible, as the angle between a few thousand km and a thousand au is very small.

The astrometric uncertainty of the probe's position in the SGL is smaller than 0.1\,arcsec, and will be dominated by the transmitter offset as explained in the next section. The distance between the Earth and the probe at $z=1{,}000\,$au is, for this epoch, 999.22\,au.

\begin{figure*}
\includegraphics[width=\linewidth]{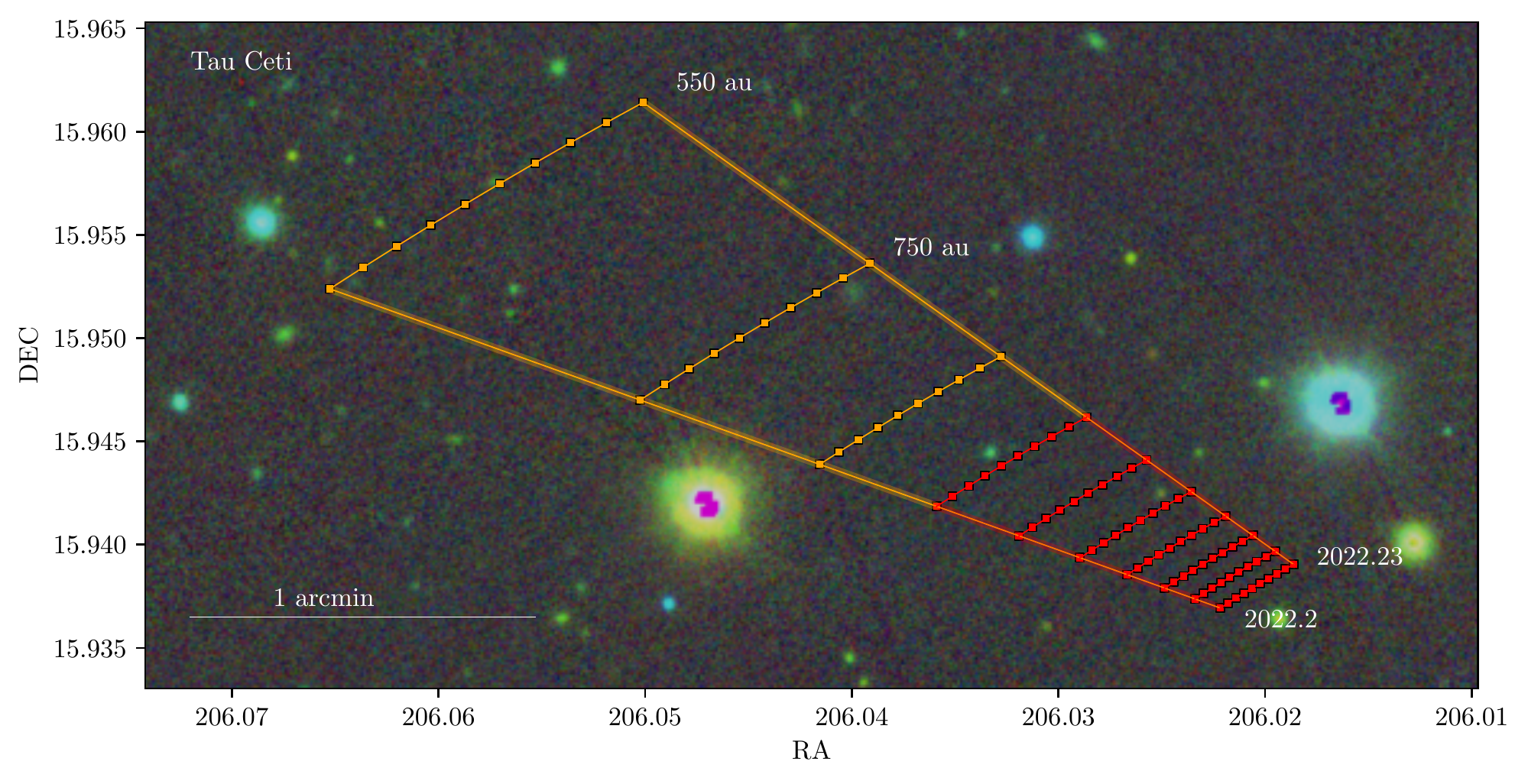}
\caption{Apparent movement of the SGL receiver. Each square symbol shows the position after 24\,hrs, for a series of distances (lines). 
Location of a probe in the sun's gravitational lens for Tau Ceti, as seen from the Earth. The two thick lines show the SGL receiver positions as a function of heliocentric distance, for 2020.2 (lower line, 13 March 2020) and 2020.3 (upper line, 23 March 2020). The apparent movement of the probe is shown with a marker for each day between these dates. The width of the lines represent the apparent location shift due to a transmitter with a 1\,au distance from Tau Ceti. This width is much smaller than the apparent sky movement due to the parallax motion due to Earth's motion around the sun. The apparent shift caused by the movement of the transmitter near Tau Ceti due to its orbit around Tau Ceti can be entirely neglected, as it is extremely small compared to the other factors shown here. Depending on heliocentric distance and the time of the year, the apparent movement of the probe is detectable during a single night, or at most after a few days, as its apparent motion is usually of order arcsec per day.}
\label{fig:tau_ceti_few_days}
\end{figure*}

\subsection{Offset of transmitter from Tau Ceti}
The largest possible deviation of a probe's position from a line of positions is due to the distance between the transmitter and its host star, causing a displacement of $\theta=1\,$arcsec per au of separation at a distance of 1\,pc; or more generally,

\begin{eqnarray}
\begin{aligned}
\theta = 1\,
\left( \frac{a}{1\,{\rm au}} \right)
\left( \frac{1\,{\rm pc}}{d} \right)
\,{\rm arcsec.}
\end{aligned}
\end{eqnarray}

For a semimajor axis of 1\,au at a distance of 1\,pc, the maximum displacement of the transmitter from Tau Ceti is $\sim0.27\,$arcsec. Compared to this uncertainty, typical errors from Gaia astrometry are smaller.

The probability that the transmitter is near maximum separation at any given time is quite high; this fact is known as the orbital sampling effect \citep{2014ApJ...787...14H}. The normalized sampling frequency as a function of phase shows that the orbiting body spends 50\,\% of its time at the largest 20\,\% of apparent separations.

\subsection{Neglected corrections}	
For our use case of determining a sky location with arcsec accuracy, it is sensible to neglect corrections which are irrelevant. There is a myriad of ever smaller effects at the sub-arcsec level. Considering these would not change the result, but only introduce potential errors. However, we will list them with their magnitude as follows, to allow for verification and implementation in relevant cases.	

It is sufficient to use $z$ values with respect to the Sun, instead of the Earth, as the error is $<2\,$au, i.e. $<20\,$min of light travel time. The resulting SGL displacement is $<0.03\,$arcsec.

A similarly small error comes from the fact that we have calculated the position of Tau Ceti with proper motion for the observation epoch as seen from the Earth, instead of the SGL. The true proper motion is different by the light travel time needed to traverse $\pm 1{,}000$au. Proper motions for Tau Ceti are $\sim2\,$arcsec per year, and $2{,}000\,$au add only 12 days to this, i.e. $<0.1\,$arcsec. Therefore, this correction can be neglected.

\subsection{Search area}
With the location and its uncertainties established, how large is the search area? For a moment in time, the probe's position is a point on the focal line, with the point's position being a function of $z$ (which we do not know). The apparent length of this line for $550<z<2{,}500\,$au is typically a few arcmin, and about one arcmin for the favored region of $1{,}000<z<2{,}500\,$au (the additional line length goes to zero when $z$ goes to infinity). The width of the line is dominated by our uncertainty in the offset between probe and star. Even for close stars, this is less than an arcsec per au. The search area per night is of order 60 square arcsec. (Figure~\ref{fig:tau_ceti_few_days}). The apparent movement of the probe's position is a few arcsec.

\begin{figure}
\includegraphics[width=\linewidth]{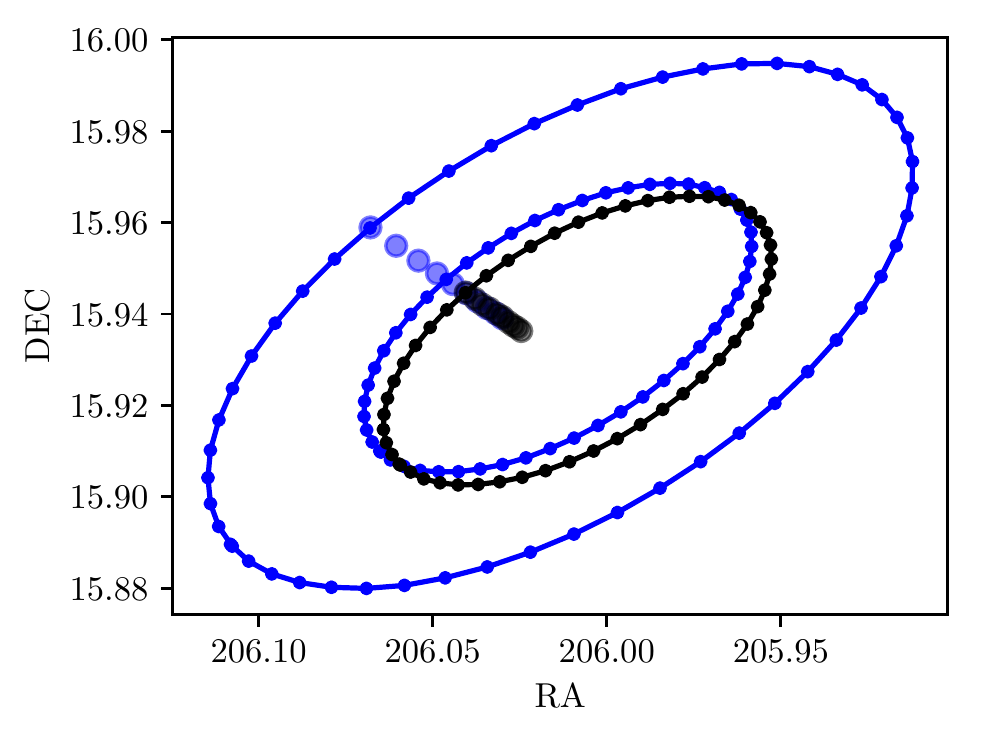}
\caption{Receiver (black) and transmitter (blue) locations over the course of 52 weeks. Small circles are for $z=1{,}000$, large circle for $550\,$au. The streak of points covers the range for $550<z<2{,}500\,$au at the epoch 2020.2. Jitter of points is due to random sampling of a 1\,au wide orbit at Tau Ceti.}
\label{fig:transmitter}
\end{figure}

\section{Localization of a transmitter node in the solar gravitational lens}
\label{sec:localization_transmitter}
We now determine the location of a transmitter $D_{\rm SGL,t}$ in the image plane, building on the framework of the receiver. A transmitter can beam photons into the Einstein ring, focusing them in a pencil beam towards Tau Ceti. If all objects were static, the transmitter location would be identical to the receiver location. Due to the proper motion of Tau Ceti, it is not. A node might serve double duty by switching positions after certain (perhaps decade-long) times, or multiple nodes may be used. Some links may be one-way, so that it is useful to check both receiver and transmitter locations for each target.

Transmitters benefit from heliocentric positions close to the minimum $z\sim 547.8\,$au, because a larger part of their flux gets injected into the Einstein ring, and they do not suffer from coronal noise like receivers. The free space receiver near Tau Ceti, however, faces the issue that it is difficult to spatially resolve an Einstein ring so close to the sun. Starlight and Einstein ring are resolved by free space receiver apertures

\begin{eqnarray}
\begin{aligned}
\nonumber
D_r > 140\,
\left(\sqrt{\frac{z}{550{\,\rm au}}} -1 \right)^{-1}
\left(\frac{\lambda}{1\,\mu{\rm m}} \right)^{-1}
\left(\frac{d}{1\,{\rm pc}}\right)
\,{\rm m}
\end{aligned}
\end{eqnarray}

which is about 500\,m for $z=1{,}000\,$au over a distance of 3.6\,pc to Tau Ceti; much larger than the small (meter-sized) receiver expected in the scheme. Thus, it is not expected that a receiver would spatially resolve the signal flux in the Einstein ring from the stellar noise. Consequently, long-time transmitters are most logically placed at short heliocentric distances. Transmitters which serve double duty as receivers must move between both positions, which is easier for large heliocentric distances, because coronal noise in the receiver scales as $z^{-3.5}$.

The apparent transmitter sky ellipse is the same as for the receiver, but shifted in RA and DEC due to the proper motion of the target star (Figure~\ref{fig:transmitter}). For Tau Ceti, this is $\sim2\,$arcsec per year, or $\sim23\,$arcsec over the light travel time. For the example epoch 2020.2, we observe the probe on the focal line towards the location where Tau Ceti was in the year $2020.2 - 11.753$ years minus 5.8\,d of light time travel (at $z=1{,}000\,$au) from the SGL to Earth. Then, we see the probe at 
${\rm RA}=206.0505$, 
${\rm DEC}=15.9445$
when observing from Earth at epoch 2020.2. The uncertainty in proper motion is $\lesssim1\,$arcsec. The uncertainty in the probe location from the error in distance (light travel time) is $\lesssim0.1\,$arcsec.

The centers of the receiver and transmitter ellipses are offset by 23\,arcsec as seen from the Earth. In the image plane, the physical separation between the two locations is 0.11\,au. This value scales with the proper motion of the target star; it is smaller for lower proper motions. The extreme case is for Barnard's star, where the separations are about ten times as large compared to Tau Ceti. For distant (kpc) target like the galactic center, they are zero for all practical purposes.

\section{Detection limits}
\label{sec:seta}
What is the apparent brightness of a probe in the SGL as a function of distance and size? If a probe sits in the SGL, can we see it?

\subsection{Leakage from the transmitter beam}
The transmitter beam width at arrival in the solar system is a most likely just a few $R_{\odot}$ for optimal Einstein ring illumination. For comparison, Earth's distance from the sun is $1\,{\rm au}\sim217\,R_{\odot}$, or $\sim 100 \sin i \times$ as far as the PSF. The beam power will likely be of order kW, and certainly below GW due to mode saturation. In total, the number of photons from leakage, at Earth's orbit around the sun, is essentially zero for any plausible configuration. Similarly, the chance alignment of Earth being in the path of a distant (pc) transmitter beaming at some other star is small, but non-negligible for signals inside the ecliptic plane \citep{2014JBIS...67..232F}.

\subsection{Reflected light from the SGL receiver}
Direct imaging in the optical poses the problem that a meter-sized object at $1{,}000\,$au distance has an apparent brightness well beyond the 31st magnitude even for fortunate reflection angles and high reflectivity \citep{2014AcAau..94..629G}. Objects so dim are undetectable even with 30\,m class telescopes. The problem is less severe for closer objects. For flat mirrors with specular reflection, the limit for LSST is about a meter-sized object at 20\,au (Uranus), which scales to km size at $1{,}000\,$au \citep{2019PASP..131h4401L}. Finding a super-Earth sized ``planet~9'' at hundreds of au is difficult \citep{2019PhR...805....1B}. Limits for LSST at $1{,}000\,$au are about $1\,R_{\oplus}$ assuming an albedo of 0.5 \citep{ 2018AJ....155..243T}.

The second common method to search for dim far-away objects are stellar occultations. In our case, they are equally impossible, as the probe has an apparent size of $\sim3\,$nas per meter, very small compared to the apparent diameter of stars ($20\,\mu$as for $1\,R_{\odot}$ at 250\,pc). The resulting (rare) dip in brightness is much less than a percent, with a duration of about a second.

\subsection{An easy search for a probe with a tiny light}
If the probe carries a beacon, at what power level can we see it? In this scenario, we assume the beacon not as isotropic but focused at Earth.

An optical continuous-wave kW power laser from a focusing meter-sized aperture in the SGL beams $10^4$ photons per second into a meter-sized mirror on Earth, comparable to a 15th magnitude star (22.5\,mag at 1\,W). Data from Gaia is complete to G$\sim20\,$mag, but sources which move a few arcmin between exposures over the course of months would not be recognized as one source but dismissed as noise. DR2 completeness limit is about 1\,arcsec\,yr$^{-1}$ \citep{2018A&A...616A...2L}. For $z=1{,}000\,$au, proper motion of probe is $\sim3\,$arcmin\,yr$^{-1}$, or $>100\times$ faster than the detection threshold. From Earth, a meter-sized telescope with a commercial CCD sensor gets to 22.5\,mag at ${\rm SNR}=7$ in a 30\,min exposures. Images taken over consecutive nights would show the parallax of a Watt power laser onboard a putative SGL probe.

If the transmission is pulsed it can deliver 10 photons per square meter per J of energy. Detection is thus possible with current OSETI experiments for pulse energies of a few J \citep{2016SPIE.9908E..10M}. Lasers for pulses with J energy and 0.6\,ns pulse width are commercially available; their sizes are typically of order $0.03\,$m$^3$. Such beacons are feasible with Earth 2020 technology; they could be part of a lurking probe, detectable with today's facilities.

\subsection{Messaging the probe}
\label{sec:meta}
If we can not see the probe, it might still be there. Can we send a message? Messaging to extraterrestrial intelligence (METI) has been criticized as ETI might be evil \citep[e.g.,][]{2016JBIS...69...31G,2019AJ....158..203M}. On the other hand, it has been argued that the risk does not apply to a probe in our own solar system, because the probe already knows about us \citep{2014AcAau..94..629G} -- would it mean any harm, it would have done so already. The remaining risk would be to trigger first contact with all its (unknown) consequences, e.g. fear and confusion. A discussion of risks versus benefits is outside the scope of this paper.

With the probe location known to order arcsec accuracy, we know where to beam photons at. All wavelengths are worth a try including classical radio -- after all, the probe might be monitoring Earth in a wide band. Coherent emission (laser light) at NIR, optical and X-ray frequencies are useful, because these are what works best for the lensed interstellar communication.

To beam optical (laser) photons with an arcsec angle, a small transmitting telescope ($D_t=0.1$\,m) is sufficient because the typical atmospheric seeing is of the same order. Such a link delivers of order one photon per Joule of energy to a $D_r=1$\,m receiver at $z=1{,}000\,$au. As an example, a commercial kW laser beams $1{,}000$ photons per second on the receiver aperture. This is about 1\,\% of the background flux from Earth (from reflected sunlight). The probe could still see the laser with a spectral filter tuned to the right wavelength, or if the laser is pulsed and the receiver is equipped with fast imaging equipment. For example, a 10\,J laser with 10\,ns pulse width \citep{2017OExpr..2521981L,2018OExpr..2632717C} beams of order 10 photons on the receiver, while the background flux from Earth is less than one photon in the same time. Thus, it is possible to signal the probe with modest equipment, and even possible to send a simple Morse-style message. But the attempt could also be viewed as an embarrassing showcase of our low technological level: barely sufficient to outshine the host planet.

\section{Discussion and conclusion}
Because of the large gain for communication, it appears likely that our own first interstellar exploration probe (perhaps to Tau Ceti) would be accompanied by a second probe in the opposite direction, to be used as the communication receiver. Perhaps this step of technological development constitutes the ``readyness'' of a civilization required to be granted permission into the galactic club: Come to the node to discover it, and be invited to join.

Current technology (Earth 2021) has sent probes such as Voyager to distances of about 150\,au over a time of 42 years. These probes had an initial power budget of $\sim500\,$W and still function to this day, including their mechanical tape recorder. Faster probes are possible, but to reach $1{,}000\,$au in less than a century is challenging for chemical propulsion; perhaps possible with a sun-diving Oberth manoeuvre \citep{2019AcAau.162..284M}. Due to light travel times of order weeks, onboard intelligence must be at some advanced level of AI. While it is difficult to estimate future technological progress, the requirements are perhaps already in reach of a Manhattan-style project; certainly they can be expected within the next century.

In the meantime, we can observe the SGL for various stars in the optical and radio domain. The first efforts to do so are ongoing and will be reported elsewhere.
\\
\\
\textit{Acknowledgments} I thank Jason Wright for useful discussions.

\software{
astropy \citep{2013A&A...558A..33A,2018AJ....156..123A},
matplotlib \citep{matplotlib:2007},
Jupyter Notebooks \citep{Kluyver:2016aa}
}

\bibliography{references}
\end{document}